\documentclass[12pt]{article}

\usepackage{amsmath}

\allowdisplaybreaks

\begin{document}

\baselineskip=18.6pt plus 0.2pt minus 0.1pt

\makeatletter
\@addtoreset{equation}{section}
\renewcommand{\theequation}{\thesection.\arabic{equation}}
\renewcommand{\thefootnote}{\fnsymbol{footnote}}

%%%%%%%%%%%%%%% Private Macros %%%%%%%%%%%%%%%%%%%%%%%%%%%%%%%%%%
%%%                          last update 2000/01/07

\newcommand{\der}[1]{\partial_{#1}}
\newcommand{\dder}[2]{\frac{\partial{#1}}{\partial{#2}}}
\newcommand{\inv}[1]{\frac{1}{#1}}
\newcommand{\bra}[1]{\langle \, #1 \, |}
\newcommand{\ket}[1]{|\, #1\, \rangle}
\newcommand{\sket}[2][3]{|\, #2\, \rangle_{#1}}
\newcommand{\tket}[1]{|\, #1\, \rangle_{123}}
\newcommand{\braoket}[3]{\langle \, #1 \,|\, #2 \,|\,  #3\, \rangle}
\newcommand{\bbbk}[4]{{}_1\langle \, #1 \,|\,{}_2\langle \, #2 \,|\,
                      {}_3\langle \, #3 \,|\,  #4\, \rangle_{123}}

\newcommand{\bbk}[3]{{}_1\langle \, #1 \,|\,{}_2\langle \, #2 \,|\,
                      #3\, \rangle_{123}}

\newcommand{\bbok}[4]{{}_1\langle \, #1 \,|\,{}_2\langle \, #2 \,|\,
                      #3 \,|\, #4\, \rangle_{123}}
\newcommand{\ph}[1]{\phantom{#1}}
\newcommand{\nn}{\nonumber}
\newcommand{\QB}{Q_{\text B}}
\newcommand{\tilQ}{\widetilde Q_{\text B}}
\newcommand{\delB}{\delta_{\text B}}
\newcommand{\arn}[2]{\alpha^{(#1)}_{#2}}
\newcommand{\brn}[2]{b^{(#1)}_{#2}}
\newcommand{\crn}[2]{c^{(#1)}_{#2}}
\newcommand{\N}[2]{N^{#1}_{#2}}
\newcommand{\X}[2]{X^{#1}_{#2}}
\newcommand{\Nsix}[2]{\bar N^{#1}_{#2}}
\newcommand{\alal}[2]{(\alpha_{#1}\cdot\alpha_{#2})}
\newcommand{\alalr}[3]{(\alpha_{#2}^{(#1)}\cdot\alpha_{#3}^{(#1)})}
\newcommand{\bc}[2]{b_{#1}c_{#2}}
\newcommand{\cb}[2]{c_{#1}b_{#2}}
\newcommand{\bcr}[3]{b_{#2}^{(#1)}c_{#2}^{(#1)}}
\newcommand{\cbr}[3]{c_{#2}^{(#1)}b_{#2}^{(#1)}}
\newcommand{\pref}[1]{(\ref{#1})}
\newcommand{\vbar}[1][]{\Bigr|_{#1}}
\newcommand{\tm}{\mbox{$\times$}}
\newcommand{\kp}{\mu}
\makeatother
%%%%%%%%%%%%%%% End of Private Macros %%%%%%%%%%%%%%%%%%%%%%%%%%

\begin{titlepage}
\title{
\hfill\parbox{4cm}
{\normalsize KUNS-1683\\{\tt hep-th/0009105}}\\
\vspace{1cm}
BRST Invariance of the Non-Perturbative Vacuum\\[5pt]
in Bosonic Open String Field Theory
}
\author{
Hiroyuki {\sc Hata}\thanks{{\tt hata@gauge.scphys.kyoto-u.ac.jp}}
{}\hspace*{5pt} and \hspace*{5pt}
Shun'ichi {\sc Shinohara}\thanks{
{\tt shunichi@gauge.scphys.kyoto-u.ac.jp}}
\\[7pt]
{\it Department of Physics, Kyoto University, Kyoto 606-8502, Japan}
}
\date{\normalsize September, 2000}
\maketitle
\thispagestyle{empty}
\begin{abstract}
\normalsize\noindent
Tachyon condensation on a bosonic D-brane was recently demonstrated
numerically in Witten's open string field theory with level truncation
approximation.
This non-perturbative vacuum, which is obtained by solving the
equation of motion, has to satisfy furthermore the requirement of
BRST invariance. This is indispensable in order for the theory around
the non-perturbative vacuum to be consistent. We carry out the
numerical analysis of the BRST invariance of the solution and find
that it holds to a good accuracy.
We also mention the zero-norm property of the solution.
The observations in this paper are expected to give clues to the
analytic expression of the vacuum solution.

\end{abstract}

\end{titlepage}

%%%%%%%%%%%%%%%%%%%%%%%%%%%%%%%%%%%%%%%%%%%%%%%%%%%%%%%%%%%%%%%%%%%%
\section{Introduction}

Recently, string field theories, especially Witten's
open string field theory \cite{Witten:1986cc}, have attracted much
interest from the viewpoint of tachyon condensation.
In the course of studies on non-BPS states in string
theory \cite{Sen:1999mg,Sen:1998sm,Sen:1999mh},
Sen has made the following conjectures.
\begin{itemize}
\item The open string tachyon is a sign of instability of
  the system, and it condenses into the non-perturbative vacuum,
  where the space-time filling D25-brane completely disappears.
\item Kink solutions on the non-perturbative vacuum represent various
  lower dimensional \mbox{D-branes} (descent relations).
\end{itemize}
To prove these conjectures, especially the first one, we need an
off-shell formulation of string theory, and a unique candidate at
present is string field theory.
In fact, Sen and Zwiebach \cite{SZ} showed that the first conjecture
holds to a miraculously good accuracy by making use of string field
theory with level truncation scheme \cite{Kostelecky:1988ta}.
After this breakthrough, a lot of works in this direction have come
out. For the first conjecture, there have appeared works on
higher order level truncations \cite{Moeller:2000xv},
superstring extension
\cite{Berkovits:2000zj,Berkovits:2000hf,DeSmet:2000dp,Iqbal:2000qg},
and approaches toward the analytic solution
\cite{Sen:1999xm,0006240,Zwiebach:2000dk,Minahan:2000ff,
Kostelecky:2000hz}.
As for the second conjecture, there are approaches in the level
truncation scheme
\cite{Harvey:2000,deMelloKoch:2000ie,Moeller:2000jy,0008053,0008101}
and the analysis in the large non-commutativity limit
\cite{0005006,Harvey:2000jt,Witten:2000nz,Gopakumar:2000rw}.
In particular, the latter has given a decisive answer to the descent
relations.

Because Witten's open string field theory is formulated as a gauge
theory, we must fix the gauge invariance to find classical solutions.
To fix the gauge we use the BRST method, and then the gauge-fixed
action has an invariance under the BRST transformation.
The BRST invariance of the vacuum as well as of the action is by all
means necessary to deal with theories with unphysical negative-norm
states \cite{Kugo:1979gm}. The perturbative vacuum with vanishing
string field is
trivially BRST-invariant. However, the BRST invariance of the
tachyon condensed vacuum is a non-trivial matter since the vacuum
solution is obtained by solving only the equation of motion.
So we must check the BRST invariance separately to guarantee that the
perturbation theory around the vacuum makes sense physically.

The purpose of the present paper is to examine the BRST invariance of
the non-perturbative vacuum solution found in \cite{SZ,Moeller:2000xv}
in the level truncation approximation.
We find that the BRST invariance for the lower level states hold to a
very good accuracy.
We also analyze the ``fake'' vacuum belonging to a different branch
from the ``true'' one and show that the BRST invariance is
spontaneously broken there.
Though we analyze the BRST invariance only numerically here,
the BRST invariance of the full theory is expected to give
crucial information for constructing the analytic expression of the
vacuum solution.

The rest of this paper is organized as follows.
In section 2, we recapitulate Witten's open string field theory
to fix our conventions.
Section 3, the main part, is devoted to the analysis of the BRST
invariance of the solution.
Lastly in section 4, we conclude our results and make some comments on
the numerical results of \cite{SZ,Moeller:2000xv}.

\section{String field theory action and BRST invariance}
\label{sec:gauge-fix}

In this section we briefly review Witten's open bosonic string field
theory \cite{Witten:1986cc}.
Especially we pay attention to gauge invariance, gauge fixing and
BRST invariance.
Since we are interested only in the translation invariant states, we
set the space-time momentum equal to zero.

\noindent\underline{\bf Action}

The gauge invariant action of Witten's string field theory is
given by\footnote{
We have factored out the space-time volume $V_{26}$ since $\Phi$ is
translation invariant. $g_{\rm o}$ is the open string coupling
constant.
}
\begin{equation}
  \label{gauge-inv-S}
  S = -\frac{V_{26}}{g_{\rm o}^2} \left( \inv{2} \int db_0\;
                    \braoket{\Phi}{\QB}{\Phi}
      + \inv{3} \int db_0^{(3)}\int db_0^{(2)}\int db_0^{(1)}
                    \bbbk{\Phi}{\Phi}{\Phi}{V_3} \right).
\end{equation}
In this paper we write string field $\ket{\Phi}$ as a vector in the
Fock space of first quantized string except ghost zero modes, $b_0$
and $c_0=\dder{}{b_0}$.
String field $\ket{\Phi}$, a ghost number $-1$ state,
is expanded in terms of $b_0$ as
\begin{equation}
  \label{unfixed_Phi}
  \ket{\Phi} = b_0 \ket{\phi} + \ket{\psi} ,
\end{equation}
where $\ket{\phi}$ and $\ket{\psi}$ carry ghost number $0$ and $-1$
respectively.
The BRST operator $\QB$ is represented as
\begin{align}
  \QB & = c_0 L + b_0 M + \tilQ,
\end{align}
with
\begin{align}
  L \;\, & = \sum_{n=1}^{\infty}\left(
  \alpha_{-n}\alpha_{n} + n c_{-n}b_{n} + n b_{-n}c_{n} \right) -1,\\
  M \; & = -2 \sum_{n=1}^{\infty} n c_{-n}c_{n},\\
  \tilQ & = \sum_{n\neq 0}c_{-n} \sum_{m=-\infty}^{\infty} \inv{2}
            \alpha_{n-m}\alpha_{m}
            + \sum_{n\neq 0, m\neq 0} \frac{m-n}{2} c_m c_n b_{-m-n}.
\end{align}
The vertex $\tket{V_3}$ representing the three-string midpoint
interaction is defined by \cite{Samuel:1986yw,Gross:1987ia}
\begin{align}
  \label{V3}
  \tket{V_3} & = \exp\left[-\sum_{r,s=1}^{3}
    \sum_{n=1}^{\infty}X^{rs}_{n0}\crn{r}{-n}\brn{s}{0}
    \right]\tket{v_3}\, ,\\
  \label{v3}
    \tket{v_3} & = \kp\exp\left[\sum_{r,s=1}^{3}
    \left( \inv{2}\sum_{n,m=1}^{\infty}
    N^{rs}_{nm}\arn{r}{-n}\cdot\arn{s}{-m}
    - \sum_{n,m=1}^{\infty}X^{rs}_{nm}\crn{r}{-n}\brn{s}{-m} \right)
  \right]\tket{0}\,.
\end{align}
Here we adopt the convention of \cite{Moeller:2000xv} and take
$\kp=3^{9/2}/2^7$.
The Neumann coefficients, $N^{rs}_{nm}, X^{rs}_{nm}$ and
$X^{rs}_{n0}$ are defined using the six-string Neumann coefficients
$\Nsix{RS}{nm}$ $(R,S=1,\dots,6)$ as
\begin{align}
  \N{rs}{nm} & = \Nsix{rs}{nm} + \Nsix{r\,(s+3)}{nm}, \\
  \X{rs}{nm} & = (-)^{r+s+1}\,n\left(\Nsix{rs}{nm}
                        - \Nsix{r\,(s+3)}{nm}\right),
\end{align}
where $\Nsix{RS}{nm}$ are given by the contour integrals as
\begin{align}
  \begin{split}
      \Nsix{RS}{nm} & = \inv{nm}\oint_{Z_R} \frac{dz}{2\pi i}
                     \oint_{Z_S} \frac{dw}{2\pi i} \inv{(z-w)^2}
                    (-)^{n(R-1)+m(S-1)}[f(z)]^{n(-)^R}
                    [f(w)]^{m(-)^S},\\
                    & \hspace*{20em} n>0, \; m>0, \\[1ex]
  \Nsix{RS}{n0} & = \inv{n}\oint_{Z_R} \frac{dz}{2\pi i}
                     \inv{z-Z_S}
               (-)^{n(R-1)}[f(z)]^{n(-)^R}, \quad\quad n>0,\\[1ex]
  f(z) & = \frac{z(z^2-3)}{3z^2-1},\qquad
  Z_{R=1,\dots,6}  = \left\{\sqrt{3}, \;\inv{\sqrt{3}},\; 0,\;
                    -\inv{\sqrt{3}},\; -\sqrt{3},\; \infty \right\}.
  \end{split}
\end{align}

\noindent\underline{\bf Gauge fixing and BRST invariance}

Since the action (\ref{gauge-inv-S}) has an invariance under the
stringy gauge transformation
\begin{equation}
  \label{gauge_trf}
  \delta_{\Lambda} \sket{\Phi}=
    \QB \sket{\Lambda} + \bbk{\Phi}{\Lambda}{V_3}
   -\bbk{\Lambda}{\Phi}{V_3},
\end{equation}
we have to fix the gauge and introduce the ghost fields to quantize
the system. This procedure is accomplished by imposing the gauge
condition (here we take the Siegel gauge),
\begin{equation}
  \label{gauge_fix}
  b_0 \ket{\Phi} = 0\quad
\Leftrightarrow\quad \ket{\psi}=0,
\end{equation}
and allowing the component fields with an arbitrary ghost number in
$\ket{\phi}$. Then the action (\ref{gauge-inv-S}) with the gauge
condition (\ref{gauge_fix}) gains, instead of the gauge invariance,
the invariance under the BRST transformation
$\delB \phi = \left(\delta S/\delta\psi\right)_{\psi=0}$.

Let us summarize the concrete expressions of the equation of motion
and the BRST transformation in the gauge-fixed string field theory.
First, the equation of motion of the gauge invariant action
(\ref{gauge-inv-S}) reads
\begin{equation}
  \label{unfixed_EOM}
  \QB^{(3)} \sket{\Phi} + \int db_0^{(2)} \int db_0^{(1)}\;
       \bbk{\Phi}{\Phi}{V_3} =0.
\end{equation}
Imposing the gauge condition (\ref{gauge_fix}), the
$b_0^{(3)}$-independent part of (\ref{unfixed_EOM}) gives the equation
of motion of the gauge-fixed system:
\begin{equation}
  \label{fixed_EOM}
  L^{(3)}\sket{\phi} + \bbk{\phi}{\phi}{v_3} = 0 .
\end{equation}
On the other hand, the $b_0^{(3)}$ part of the left-hand-side of
(\ref{unfixed_EOM}) is the BRST transformation:
\begin{equation}
  \delB \sket{\phi} = \tilQ^{(3)} \sket{\phi} -
       \sum_{r=1}^{3}\sum_{n=1}^{\infty}
         \X{r3}{n0} \; \bbok{\phi}{\phi}{c_{-n}^{(r)}}{v_3}.
       \label{BRST_trf}
\end{equation}

The equation of motion (\ref{fixed_EOM}) was studied numerically in
\cite{SZ,Moeller:2000xv} using the level truncation scheme, and they
found a surprisingly good agreement with the conjectured result
\cite{Sen:1999mh} for the tachyon potential.  So the tachyon-condensed
vacuum $\phi_{\text c}$ calculated there is believed to be the
true vacuum in which D25-brane completely disappears and there are no
degrees of freedom of open strings.

However, besides being a solution to the equation of motion
(\ref{fixed_EOM}), $\phi_{\text c}$ should be BRST invariant,
$\delB \ket{\phi_{\text c}}=0$ (namely, the $b_0^{(3)}$ part of the
original equation of motion (\ref{unfixed_EOM}) should also hold).
This requirement of the unbroken BRST invariance is independent of the
equation of motion (\ref{fixed_EOM}).
Moreover, it is necessary in order for the
perturbation theory around $\phi=\phi_{\text c}$ to be a consistent
one where unphysical negative-norm states are controlled by the BRST
symmetry. The BRST invariance should, in particular, play an important
role in showing
that there are no open string excitations on the non-perturbative
vacuum. Poles in the propagators are allowed if they correspond to
unphysical states confined by the BRST symmetry \cite{Kugo:1979gm}.

Therefore, it is an important and non-trivial matter to check the BRST
invariance of the numerical solution obtained in
\cite{SZ,Moeller:2000xv}. We shall carry out this task in the next
section.

\section{BRST invariance of the non-perturbative vacuum}
\label{sec:BRST-inv}

In this section we examine the BRST invariance of the
non-perturbative vacuum of \cite{Moeller:2000xv} for the first few
levels in $\delB \ket{\phi_{\text c}}$.
We shall find more and more precise cancellations among the terms in
$\delB \ket{\phi_{\text c}}$ as we incorporate higher level fields in
the quadratic term in (\ref{BRST_trf}).

\subsection{BRST invariance at level two}
\label{sec:c-2}

To begin with, we expand the string field following
\cite{Moeller:2000xv} up to level four,
assuming that only zero-ghost-number, Lorentz-scalar and even-level
fields acquire non-zero vacuum expectation values\footnote{
We can further truncate matter CFT states to
the descendant states of unit operator \cite{Sen:1999xm}.
But we do not take this property into account in this paper.
}:
\begin{align}
  \begin{split}
    \ket{\phi} = & \ph{+i}\psi_1 \;\ket{0}
                 + \psi_2 \;\alal{-1}{-1}\ket{0}
                 + \psi_3 \;\bc{-1}{-1}\ket{0}\\
               & + \psi_4 \;\alal{-1}{-3}\ket{0}
                 + \psi_5 \;\alal{-2}{-2}\ket{0}
                 + \psi_6 \;\alal{-1}{-1}\alal{-1}{-1}\ket{0} \\
               & + \psi_7 \;\alal{-1}{-1}\bc{-1}{-1}\ket{0}
                 + \psi_8 \;\bc{-1}{-3}\ket{0}
                 + \psi_9 \;\bc{-2}{-2}\ket{0}
                 + \psi_{10} \;\bc{-3}{-1}\ket{0}.
   \end{split}
\end{align}
Because the BRST transformation $\delB$ raises the ghost number by
one, we should examine the
BRST transformation of the component fields with ghost number $-1$.
Explicitly, we shall consider the following five component fields
in $\ket{\phi}$ with levels 2 and 4:
\begin{equation}
  \begin{split}
      \label{barC}
    &\bar C_2\,c_{-2}\ket{0}
    +\bar C_{4a}\,c_{-4}\ket{0}
    +\bar C_{4b}\,b_{-1}c_{-1}c_{-2}\ket{0} \\
    &\quad +\bar C_{4c}\,\alal{-1}{-1}c_{-2}\ket{0}
    +\bar C_{4d}\,\alal{-1}{-2}c_{-1}\ket{0} .
  \end{split}
\end{equation}

First let us consider $\delB\bar C_2$.
As an example of the calculation of the quadratic term in $\delB
\ket{\phi}$ of (\ref{BRST_trf}) contributing to  $\delB\bar C_2$,
we present the level-0\tm level-0 and
level-0\tm level-2 terms using the Neumann coefficients:
\begin{align}
  \Big[ -\X{33}{20} \;(\psi_1)^2  +
   2\cdot 26  \N{11}{11}\X{33}{20} \psi_1\psi_2
  -2 \left( \X{11}{11}\X{33}{20} - \X{31}{21}\X{13}{10}\right)
      \psi_1\psi_3 \Big]\; c_{-2}\ket{0}.
\end{align}
Collecting the linear term, $\tilQ\ket{\phi}$ of (\ref{BRST_trf}),
and all the quadratic terms with the level sum equal
to or less than four, we get the following expression for
$\delB\bar C_2$:
\begin{equation}
  \label{BRS_22}
  \begin{split}
   \delB \bar C_2 & = 26\,{\psi_2} + 3\,{\psi_3}  \\
   & \quad - \kp \Biggl[  \frac{16\,{{\psi_1}}^2}{27}
   - \frac{4160\,{\psi_1}\,{\psi_2}}{729}
    + \frac{32\,{\psi_1}\,{\psi_3}}{81}
    + \frac{483392\,{{\psi_2}}^2}{19683}
   - \frac{4160\,{\psi_2}\,{\psi_3}}{2187} -
   \frac{11248\,{{\psi_3}}^2}{19683}\\
   & \hspace*{5em} +
   \frac{26624\,{\psi_1}\,{\psi_4}}{6561} +
   \frac{21632\,{\psi_1}\,{\psi_5}}{6561} +
   \frac{582400\,{\psi_1}\,{\psi_6}}{19683} -
   \frac{4160\,{\psi_1}\,{\psi_7}}{2187}\\
   & \hspace*{5em} + \frac{9856\,{\psi_1}\,{\psi_8}}{19683} +
   \frac{800\,{\psi_1}\,{\psi_9}}{2187}
   - \frac{3200\,{\psi_1}\,{\psi_{10}}}{19683} \Biggr].
  \end{split}
\end{equation}

Now let us substitute into (\ref{BRS_22}) the numerical values of
$\psi_k$ for the non-perturbative vacuum obtained in
\cite{Moeller:2000xv}.
The values of  $\psi_k$ in the (10,20) approximation of
\cite{Moeller:2000xv} is listed in Table \ref{tab:vev}.
\begin{table}[tbp]
  \begin{center}
    \leavevmode
    \begin{tabular}{lr||lr||lr}\hline
$\psi_1$ &   1.09259  & $\psi_4$ & $-0.01148$ & $\psi_7$    & 0.00786 \\
$\psi_2$ &   0.05723  & $\psi_5$ & $-0.00509$ & $\psi_8$    & 0.11654 \\
$\psi_3$ & $-0.43373$ & $\psi_6$ & $-0.00032$ & $\psi_9$    & 0.06990 \\
         &            &          &            & $\psi_{10}$ & 0.03885 \\
     \hline
    \end{tabular}
    \caption{The classical solution $\psi_k$ in the (10,20)
      approximation}
    \label{tab:vev}
  \end{center}
\end{table}
Table~\ref{tab:cancel} summarizes the result after substituting these
values into the BRST transformation (\ref{BRS_22}).
For example, the row [linear\,:\,$\psi_3$] is the value of the
second term of (\ref{BRS_22}), and the row
[quadratic\,:\,level-0\tm level-2]
is the value of the sum of the $\psi_1\psi_2$ and the $\psi_1\psi_3$
terms including the factor $-\kp$.
\begin{table}[tbp]
  \begin{center}
    \leavevmode
    \begin{tabular}{c|c||r|r|r|r} \hline
  lin. or quad. & term & raw value & partial sum 1 & partial sum 2 &
   total sum\\ \hline \hline
   linear & $\psi_2$ & 1.48798    & 0.18682 &  &
                                                \\ \cline{2-3}
   & $\psi_3$ & $-1.30116$ &         & $0.00774$\\ \cline{1-4}
            & level-0\tm level-0 & $-0.77537$ & $-0.17908$  &
               & $-0.00216$ \\ \cline{2-3}
  quadratic & level-0\tm level-2 &  0.59629  &  & \\ \cline{2-5}
            & level-2\tm level-2 & $-0.02209$ &
                         $-0.00990$ & $-0.00990$\\ \cline{2-3}
            & level-0\tm level-4 &   0.01219  & & & \\ \hline
    \end{tabular}
    \caption{The values of the terms in $\delB\bar C_2$ and their
   cancellations}
    \label{tab:cancel}
  \end{center}
\end{table}

{}From Table \ref{tab:cancel} we find that the cancellations occur every
time we sum up higher level contributions.
First, each term in the linear part of (\ref{BRS_22}) is of order
$10^0$, and they cancel each other to become $0.18682$, of order
$10^{-1}$.
The field $\psi_2$ is constructed by the matter oscillators $\alpha$,
and $\psi_3$ by the ghost ones, $b$ and $c$.
So it can be considered as a kind of matter-ghost cancellation.
Next adding to the linear part the level-0\tm level-0 and
level-0\tm level-2 contributions in the quadratic part gives a value
of order $10^{-2}$. Finally adding the contributions from
level-2\tm level-2 and level-0\tm level-4, $\delB\bar C_2$ becomes of
order $10^{-3}$.\footnote{
This should be compared with the deviation of the potential
height from the expected value,
which is of order $10^{-3}$ in the (10,20) approximation
\cite{Moeller:2000xv}.
}
Thus we have observed a marvelous cancelation among
various terms in $\delB\bar C_2$, in particular, between the linear
and the quadratic terms.
This analysis gives a strong evidence for $\delB\bar C_2=0$ at the
non-perturbative vacuum.

\subsection{Fake vacuum}
\label{sec:fake}

In the level truncation scheme there generally appear more than one
non-trivial solutions for the equations of motion, for example,
there are two solutions in the (2,6) approximation.
The non-perturbative vacuum obtained in \cite{SZ,Moeller:2000xv} was
chosen by the condition that it lies on the same branch as the trivial
solution $\phi=0$.

In this subsection, let us take the (2,6) approximation and
study the BRST invariance of the ``fake'' vacuum on a different
branch.
The potential in the (2,6) approximation is
\begin{align}
V = & -\frac{1}{2}{\psi_1}^2 + 26\,{\psi_2}^2
      - \frac{1}{2}{\psi_3}^2\nn\\
     & \quad  +
    \frac{\kp}{3}\Biggl[ {\psi_1}^3 -
     \frac{2178904}{6561}{\psi_2}^3
     - {\psi_1}^2\,
      \left( \frac{130}{9}\psi_2
        + \frac{11}{9}\psi_3 \right)
     - \frac{332332}{6561}{\psi_2}^2\,\psi_3\\
     & \hspace*{5em}  - \frac{2470}{2187}\psi_2\,{\psi_3}^2
     - \frac{1}{81}{\psi_3}^3
     + \psi_1\,\left( \frac{30212}{243}{\psi_2}^2
                  + \frac{2860\,\psi_2\,\psi_3}{243}
                  + \frac{19}{81}{\psi_3}^2 \right)
     \Biggr] . \nn
\end{align}
This potential has two non-trivial extrema\footnote{
Because all the coefficient fields
  are real, we discard complex solutions.},
which we call ``true'' vacuum  and ``fake'' one respectively
(see Table \ref{tab:T_F}).
\begin{table}[hbt]
  \begin{center}
    \leavevmode
    \begin{tabular}{c||r|r|r}\hline
       &  $\psi_1\ph{aai}$ & $\psi_2\ph{aai}$
        & $\psi_3\ph{aai}$ \\ \hline
     ``true'' &  $1.0884$  &  $0.05596$ &  $-0.3804$ \\
     ``fake'' & $-17.8041$ & $-1.68453$ & $-28.6253$ \\ \hline
    \end{tabular}
    \caption{Field values at the two extrema}
    \label{tab:T_F}
  \end{center}
\end{table}
Table \ref{tab:BRST_T_F} shows the values of $\delB\bar C_2$ for these
two vacua.
\begin{table}[htb]
  \begin{center}
    \leavevmode
    \begin{tabular}{c|c|c||r|r|r}\hline
    & lin. or quad. & term & raw value & partial sum &
                                total sum\\ \hline \hline
 T  &linear & $\psi_2$ & 1.45506    & 0.31391 &  \\ \cline{3-4}
 R  &         & $\psi_3$ & $-1.14114$ & & 0.10472  \\ \cline{2-5}
 U  &quadratic & level-0\tm level-0 & $-0.76944$ & $-0.20919$ &
                    \\ \cline{3-4}
 E &            & level-0\tm level-2 &  0.56025
   & \\ \hline\hline
 F&  linear & $\psi_2$ & $-43.798$  & $-129.674$ &
                                                \\ \cline{3-4}
 A&          & $\psi_3$ & $-85.876$ &  & $-368.661$ \\ \cline{2-5}
 K&quadratic & level-0\tm level-0 & $-205.889$ & $-238.987$ &
               \\ \cline{3-4}
 E&  & level-0\tm level-2 & $-33.098$
  & \\ \hline
    \end{tabular}
    \caption{$\delB\bar C_2$ for the ``true'' and ``fake'' vacua}
    \label{tab:BRST_T_F}
  \end{center}
\end{table}
In the case of the ``true'' vacuum, there occur neat cancellations as
before and total sum is close to zero.
In contrast, for the ``fake'' vacuum, all values are of order $10^2$
and there are no significant cancellations at all.
{}From these facts, we see that we can distinguish the true vacuum out
of other extrema of the potential by requiring the BRST invariance.

\subsection{BRST invariance at level four}
\label{sec:level4}

The BRST transformation of the level-4 fields in (\ref{barC}),
$\bar C_{4a}$, $\bar C_{4b}$, $\bar C_{4c}$ and $\bar C_{4d}$,
are given as follows:
\begin{align}
  \begin{split}
    \label{dBC4a}
    \delB \bar C_{4a} &=
  78\,\psi_4 + 104\,\psi_5 +
   5\,\psi_8 + 6\,\psi_9 + 7\,\psi_{10}\\
     & \qquad- \kp\Biggl[
      -\frac{352\,{\psi_1}^2}{729}
      +\frac{91520\,\psi_1\,\psi_2}{19683}
      -\frac{2240\,\psi_1\,\psi_3}{19683}
      -\frac{10634624\,{\psi_2}^2}{531441}
      +\frac{291200\,\psi_2\,\psi_3}{531441}\\
     & \hspace*{5em} +\frac{18848\,{\psi_3}^2}{59049}
      -\frac{585728\,\psi_1\,\psi_4}{177147}
      -\frac{475904\,\psi_1\,\psi_5}{177147}
      -\frac{12812800\,\psi_1\,\psi_6}{531441}\\[5pt]
     & \hspace*{5em} +\frac{291200\,\psi_1\,\psi_7}{531441}
     +\frac{262400\,\psi_1\,\psi_8}{531441}
      -\frac{27328\,\psi_1\,\psi_9}{531441}
      +\frac{256\,\psi_1\,\psi_{10}}{177147}
 \Biggr],
\end{split}
\\
\begin{split}
  \label{dBC4b}
  \delB \bar C_{4b} &=
   26\,\psi_7 - \psi_8 - 3\,\psi_9 + 5\,\psi_{10}\\
     & \qquad - \kp\Biggl[
      \frac{176\,{\psi_1}^2}{729}
     -\frac{45760\,\psi_1\,\psi_2}{19683}
     +\frac{5216\,\psi_1\,\psi_3}{19683}
     +\frac{5317312\,{\psi_2}^2}{531441}
     -\frac{678080\,\psi_2\,\psi_3}{531441}\\
     & \hspace*{5em}-\frac{592\,{\psi_3}^2}{6561}
     +\frac{292864\,\psi_1\,\psi_4}{177147}
     +\frac{237952\,\psi_1\,\psi_5}{177147}
     +\frac{6406400\,\psi_1\,\psi_6}{531441}\\[5pt]
     & \hspace*{5em}-\frac{678080\,\psi_1\,\psi_7}{531441}
     +\frac{13184\,\psi_1\,\psi_8}{59049}
     +\frac{181600\,\psi_1\,\psi_9}{531441}
     -\frac{4480\,\psi_1\,\psi_{10}}{531441}
\Biggr],
\end{split}
\\
\begin{split}
  \label{dBC4c}
  \delB \bar C_{4c} &=
  3\,\psi_4 + 56\,\psi_6 + 3\,\psi_7 + \frac{1}{2}\,\psi_9\\
    & \qquad- \kp\Biggl[
     - \frac{40\,{\psi_1}^2}{729}
     +\frac{18592\,\psi_1\,\psi_2}{19683}
     -\frac{80\,\psi_1\,\psi_3}{2187}
     -\frac{670432\,{\psi_2}^2}{177147}
     +\frac{18592\,\psi_2\,\psi_3}{59049}\\
    & \hspace*{5em}+\frac{28120\,{\psi_3}^2}{531441}
     -\frac{74752\,\psi_1\,\psi_4}{177147}
     -\frac{31168\,\psi_1\,\psi_5}{531441}
     -\frac{138880\,\psi_1\,\psi_6}{19683}\\[5pt]
    & \hspace*{5em}+\frac{18592\,\psi_1\,\psi_7}{59049}
     -\frac{24640\,\psi_1\,\psi_8}{531441}
     -\frac{2000\,\psi_1\,\psi_9}{59049}
     +\frac{8000\,\psi_1\,\psi_{10}}{531441}
\Biggr],
\end{split}
\\[7pt]
\label{dBC4d}
\delB \bar C_{4d} &=
  3\,\psi_4 + 4\,\psi_5 + 2\,\psi_7 + \psi_{10}
 - \kp\left[-\frac{65536\,\psi_2\,\psi_3}{531441}
  + \frac{65536\,\psi_1\,\psi_7}{531441} \right].
\end{align}
In Tables \ref{tab:c-4} -- \ref{tab:alalc1-4},
we list the numerical values of eqs.\ (\ref{dBC4a}) -- (\ref{dBC4d})
for the ``true'' vacuum of
Table \ref{tab:vev}.
In every case, there occur significant cancellations.
\begin{table}[htbp]
  \begin{center}
    \leavevmode
    \begin{tabular}{c|c||r|r|r}\hline
  lin. or quad. & term & raw value & partial sum &
                           total sum\\ \hline \hline
   linear & $\psi_4,\psi_5 $ & $-1.42480$ & $-0.15075$ &
                                           \\ \cline{2-3}
    & $\psi_8,\psi_9,\psi_{10}$ & $1.27405$ &  \\ \cline{1-4}
            & level-0\tm level-0 & $0.63178$ & & $-0.01672$
               \\ \cline{2-3}
  quadratic & level-0\tm level-2 & $-0.37778$
            & 0.13403& \\ \cline{2-3}
            & level-2\tm level-2 &   0.02093  &  \\ \cline{2-3}
            & level-0\tm level-4 & $-0.14090$ & & \\ \hline
    \end{tabular}
    \caption{$\delB\bar C_{4a}$}
    \label{tab:c-4}
  \end{center}
\end{table}
\begin{table}[htbp]
  \begin{center}
    \leavevmode
    \begin{tabular}{c|c||r|r|r}\hline
  lin. or quad. & term & raw value & partial sum &
                                 total sum\\ \hline \hline
   linear & $\psi_7,\psi_{10} $ & $0.39861$ & $0.07237$ &
                                                \\ \cline{2-3}
            & $\psi_8,\psi_9$ & $-0.32624$ &  \\ \cline{1-4}
            & level-0\tm level-0 & $-0.31589$ & & $-0.01040$
               \\ \cline{2-3}
  quadratic & level-0\tm level-2 &  $0.29698$
            & $-0.08277$ & \\ \cline{2-3}
            & level-2\tm level-2 & $-0.05203$  &  \\ \cline{2-3}
            & level-0\tm level-4 & $-0.01183$ & & \\ \hline
    \end{tabular}
    \caption{$\delB\bar C_{4b}$}
    \label{tab:bcc-4}
  \end{center}
\end{table}
\begin{table}[htbp]
  \begin{center}
    \leavevmode
    \begin{tabular}{c|c||r|r|r}\hline
  lin. or quad. & term & raw value & partial sum &
                            total sum\\ \hline \hline
   linear & $\psi_4,\psi_6 $ & $-0.05236$ & $0.00617$ &
                                           \\ \cline{2-3}
            & $\psi_7,\psi_9$ & $0.05853$ &  \\ \cline{1-4}
            & level-0\tm level-0 & $0.07179$ & & $0.00225$
               \\ \cline{2-3}
  quadratic & level-0\tm level-2 & $-0.08374$
            & $-0.00392$ & \\ \cline{2-3}
            & level-2\tm level-2 &   0.01124 &  \\ \cline{2-3}
            & level-0\tm level-4 & $-0.00322$ & & \\ \hline
    \end{tabular}
    \caption{$\delB\bar C_{4c}$}
    \label{tab:alalc2-4}
  \end{center}
\end{table}
\begin{table}[htbp]
  \begin{center}
    \leavevmode
    \begin{tabular}{c|c||r|r|r}\hline
  lin. or quad. & term & raw value & partial sum &
                               total sum\\ \hline \hline
   linear & $\psi_4,\psi_5 $ & $-0.05480$ & $-0.00023$ &
                                       \\ \cline{2-3}
            & $\psi_7,\psi_{10}$
            & $0.05457$ &  \\ \cline{1-4}
            & level-0\tm level-0 & $0$ & & $-0.00475$
               \\ \cline{2-3}
  quadratic & level-0\tm level-2 & $0$
            & $-0.00452$    & \\ \cline{2-3}
            & level-2\tm level-2 & $-0.00336$  &  \\ \cline{2-3}
            & level-0\tm level-4 & $-0.00116$ & & \\ \hline
    \end{tabular}
    \caption{$\delB\bar C_{4d}$}
    \label{tab:alalc1-4}
  \end{center}
\end{table}

\section{Summary and discussions}
\label{sec:conc.}

In this paper, we have shown that the non-perturbative vacuum solution
obtained in \cite{SZ,Moeller:2000xv} is BRST-invariant.
This property is necessary in order that there exists a consistent
quantum theory of fluctuations around this vacuum.
In particular, the BRST invariance would play an important role in
showing the absence of physical open string excitations on the
non-perturbative vacuum.
We have also examined the BRST invariance of the ``fake'' vacuum and
found that the BRST symmetry is spontaneously broken there.
The fact that ``fake'' vacua are not BRST-invariant may indicate that
they are artifacts of the  level truncation and do not exist in the
full theory.

Now we shall comment on the zero-norm property of the non-perturbative
vacuum solution in the level truncation scheme.
As mentioned in \cite{SZ}, if we parameterize the level-2 part of the
solution as
\begin{equation}
\label{uv}
v\cdot\inv{\sqrt{52}}\alal{-1}{-1}\ket{0} - u \cdot\bc{-1}{-1}\ket{0},
\end{equation}
the coefficients $u$ and $v$ are almost equal\footnote{
  In terms of $\psi_2$ and $\psi_3$, this
  relation reads $\psi_2 = -(1/\sqrt{52})\,\psi_3$. The BRST
  invariance,
  $\delB\bar C_2=0$, at the linearized level requires
  $\psi_2 =-(3/26)\,\psi_3$ (see eq.\ (\ref{BRS_22})).
  These two are close numerically,
  $1/\sqrt{52} = 0.139 \simeq 0.115 = 3/26$.
  The zero-norm property may possibly be related to the BRST
  invariance.
}.
Once we assume that this equality is exact, the level-2 part of the
solution has zero-norm: the norm of the first term of (\ref{uv}) is
cancelled by that of the second.
Next let us proceed to level 4 and observe that the
  $\psi_4,\psi_8$ and $\psi_{10}$ part of the solution
is rewritten
using the values of the (10,20) approximation in Table \ref{tab:vev}
as\footnote{
The classical solution $\ket{\phi_{\text c}}$ has an exact symmetry
under the exchange $(n c_{-n},b_{-m})\to (b_{-n},-m c_{-m})$,
which is owing to the invariance of $L$ and $\ket{v_3}$ under the
exchange.
We have used this fact for the second term of (\ref{zero_norm13}).
}
\begin{equation}
  \label{zero_norm13}
  -0.1014 \tm \inv{\sqrt{3\cdot 26}}\alal{-1}{-3}
  \;\;+\;\;0.0955 \tm \inv{\sqrt{6}}\bigl(3\,\bc{-1}{-3}
         + \bc{-3}{-1} \bigr),
\end{equation}
where each state is normalized to a positive or negative unit norm.
Similarly to the previous case (\ref{uv}), the two coefficients are
almost equal in a few percent error:
\begin{equation}
  \label{evaluate_error}
  \frac{(-0.1014)^2 - (0.0955)^2}{(-0.1014)^2 + (0.0955)^2}
     = 0.0599 \sim 6.0 \% .
\end{equation}
Though, in eqs.\ (\ref{zero_norm13}) and (\ref{evaluate_error}),
we considered a part of the level-4 states orthogonal to the
other parts, the zero-norm property holds better for the whole of the
level-4 state.
Evaluating the errors in the same manner as (\ref{evaluate_error}),
the whole level-4 state deviates from the zero-norm combination only
by $2.6 \%$, and the whole level-6 state by $1.5 \%$.
Therefore, it is very likely that in each level the
whole state has zero norm in the full solution (or the zero norm
property might hold in a more severe sense:
each level is a sum of zero norm states, such as (\ref{zero_norm13}),
which are orthogonal to one another).
If this is the case, we can write the potential at the extremum
normalized by the D25-brane tension $2/\pi^2$ as
\begin{align}
  \frac{\pi^2}{2}V &
     =  \frac{\pi^2}{2} \left[\frac{1}{2}\braoket{\phi}{L}{\phi}
      + \inv{3}\bbbk{\phi}{\phi}{\phi}{v_3}\right] \nn\\
    & = \frac{\pi^2}{12}\braoket{\phi}{L}{\phi}
      = -\frac{\pi^2}{12} (\psi_1)^2 ,
  \label{pot_by_tachyon}
\end{align}
where we have used the equation of motion (\ref{fixed_EOM}) at the
second equality, and used the zero-norm assumption at the last
equality.
Note that the last expression is given only in terms of $\psi_1$.
The numerical value of the last expression of (\ref{pot_by_tachyon})
is $-0.9818$, very close to the desired value $-1$.
This (almost) zero-norm property as well as the BRST invariance may
serve as important clues to the construction of the analytic solution
for the non-perturbative vacuum.

\section*{Acknowledgments}
We would like to thank T.\ Asakawa, S.\ Kawamoto and T.\ Kugo for
valuable discussions and useful comments.
This work is supported in part by Grant-in-Aid for Scientific Research
from Ministry of Education, Science, Sports and Culture of Japan
(\#12640264 and \#03529).
The work of S.\ S.\ is supported in part by the Japan Society for the
Promotion of Science under the Predoctoral Research Program.

\end{document}